%% file: main.tex
\documentclass[aps, prb, amsmath, twocolumn, longbibliography, superscriptaddress, footinbib, 10pt, final]{revtex4-2}

\makeatletter
\def\frontmatter@maketitle{%
  \@author@finish
  \title@column\titleblock@produce
  \suppressfloats[t]%
}%
\makeatother

\newcommand\includeSI{0}
\usepackage{ifthen}
\usepackage{float}
\usepackage{mathtools}
\usepackage{booktabs}
\usepackage[version=4]{mhchem} 
\usepackage{physics}
\usepackage{amssymb, amsmath}
\usepackage{hyphenat}
\usepackage{comment}
\usepackage{graphicx} 
\usepackage{epstopdf}  
\usepackage{blindtext}
\usepackage{siunitx} 
\usepackage[final]{microtype} 
\usepackage{setspace}
\usepackage[T1]{fontenc}
\usepackage{lmodern}
\usepackage{xcolor}
\usepackage{multirow}
\usepackage[colorlinks=true,linkcolor=black, citecolor=blue, urlcolor=blue]{hyperref}
\usepackage[utf8]{inputenc}
\usepackage[acronym, shortcuts]{glossaries}
\usepackage[l2tabu, orthodox]{nag}
\usepackage{makecell}
\usepackage{color,soul}
\usepackage{adjustbox}
\usepackage{environ} 
\DeclareUnicodeCharacter{2212}{-}

\sethlcolor{white}

\begin{document}

\input{customCommands.tex}

\input{acronymDefinition.tex}

\title{Enhanced thermal stability of \SiGeSn{}{}{} by suppressing surface\hyp{}mediated degradation}

\author{Anis Attiaoui} 
\altaffiliation[Present address: ]{Department of Materials Science and Engineering, Stanford University, Stanford, California 94305, USA}

\affiliation{\polydept{}}

\author{Sebastian Koelling} 
\affiliation{\polydept{}}

\author{Lu Luo} 
\affiliation{\polydept{}}



\author{Simone Assali} 
\affiliation{\polydept{}}

\author{Oussama Moutanabbir}
\affiliation{\polydept{}}

\input{0_abstract.tex}
\maketitle

\input{1_introduction_APL.tex}

\input{2_body_APL.tex}
\input{3_conclusion.tex}


\input{acknowledgments.tex}

\medskip
\noindent {\textbf{AUTHORS DECLARATIONS}}.\\
\noindent {\textbf{Conflict of Interest}}.
The authors have no conflicts to disclose.

\medskip
\noindent {\textbf{Authors Information}}.
Correspondence and requests for materials should be addressed to~:\\
\href{mailto:anatt@stanford.edu}{anatt@stanford.edu}\\
\href{mailto:oussama.moutanabbir@polymtl.ca}{oussama.moutanabbir@polymtl.ca}\\

\medskip
\noindent {\textbf{Author Contributions}}.\\
\textbf{A. Attiaoui}: Conceptualization (lead), Data curation (lead); Investigation (lead); Methodology (equal); Visualization (equal); Writing – original draft (lead); Writing – review \& editing (equal). \textbf{S. Koelling}: Data curation (equal); Formal analysis (supporting); Investigation (equal); Methodology (equal); Visualization (equal); Writing – original draft (supporting); Writing – review \& editing (equal).\textbf{ S. Assali}: Investigation (supporting); Methodology (equal); Writing – review \& editing (supporting). \textbf{L. Luo}: Resources (supporting). \textbf{O. Moutanabbir}: Conceptualization (lead); Funding acquisition (lead); Writing – original draft (equal); Writing – review \& editing (equal)

\medskip
\noindent {\textbf{DATA AVAILABILITY}}.
\par The data supporting the findings of this study are available from the corresponding authors upon reasonable request.

\bigskip

\let\oldaddcontentsline\addcontentsline
\renewcommand{\addcontentsline}[3]{}

\bibliography{references_sigesn_main.bib} 
\bibliographystyle{apsrev4-2} 

\end{document}

%% file: customCommands.tex
\def\figureautorefname{Fig.} 
\def\tableautorefname{Table} 
\newcommand{\fref}[2]{\autoref{#1}\textcolor{blue}{#2}}

\newcommand{\paperSection}[2][normal]{
    \ifthenelse{\equal{#1}{normal}}{
        \medskip\noindent{\textbf{#2}}\newline
    }{
        \noindent\textbf{#2}\newline
    }
}

 
\newcommand{\RomanNumeralCaps}[1]
    {\MakeUppercase{\romannumeral #1}}

\newcommand{\polydept}{Department of Engineering Physics, \'Ecole Polytechnique de Montr\'eal, C.P. 6079, Succ. Centre-Ville, Montr\'eal, Qu\'ebec, Canada H3C 3A7}

\newcommand{\citechange}[1]{\textcolor{red}{#1}}
\newcommand{\kp}{\bm{k \cdot p}}  
\newcommand{\GeSn}[2]{Ge$_{#1}$Sn$_{#2}$}
\newcommand{\SiGeSn}[3]{Si$_{#1}$Ge$_{#2}$Sn$_{#3}$}
\newcommand{\SipGeSn}[3]{(Si)$_{#1}$Ge$_{#2}$Sn$_{#3}$}
\newcommand{\SiGe}[2]{Si$_{#1}$Ge$_{#2}$}
\newcommand{\quot}[1]{``#1''} 
\newcommand{\acc}[1]{\Gls*{#1}}
\newcommand{\RNum}[1]{\uppercase\expandafter{\romannumeral #1\relax}}

\newcommand{\citesupp}[1]{
    \footnote{See Supplemental Material at \textbf{[URL will be inserted by publisher]} for \textcolor{red}{details of the experimental setup}, which includes Refs. #1}
}


\newcommand{\figCiteChange}[1]{\colorbox{yellow!50!white}{#1}}
\DeclareSIUnit{\million}{\text{million}}


\newcommand\vertarrowbox[3][3ex]{%
  \begin{array}[t]{@{}c@{}} #2 \vspace{1ex}\\
  \left\uparrow\vcenter{\hrule height #1}\right.\kern-\nulldelimiterspace\\
  \makebox[0pt]{#3}
  \end{array}%
}

\NewEnviron{myequation}[1]{%
\begin{equation}
\scalebox{#1}{$\BODY$}
\end{equation}
}

\newcommand{\suppfref}[2]{\autoref{#1}\textcolor{blue}{#2}}

\newcommand\suppFigRef[3]{%
    \ifthenelse{\equal{\includeSI}{1}}{%
        \let\tempfigureautorefname\figureautorefname%
        \renewcommand\figureautorefname{Fig.}%
        \suppfref{#1}{#2}%
        \let\figureautorefname\tempfigureautorefname%
    }{\textcolor{blue}{Fig. S#3}}%
}

\newcommand\suppFigsRef[3]{%
    \ifthenelse{\equal{\includeSI}{1}}{%
        \let\tempfigureautorefname\figureautorefname%
        \renewcommand\figureautorefname{Figs.}%
        \suppfref{#1}{#2}%
        \let\figureautorefname\tempfigureautorefname%
    }{\textcolor{blue}{Figs. S#3}}%
}

\newcommand\suppTabRef[3]{%
    \ifthenelse{\equal{\includeSI}{1}}{%
        \let\temptableautorefname\tableautorefname%
        \renewcommand\tableautorefname{Table }%
        \suppfref{#1}{#2}%
        \let\tableautorefname\temptableautorefname%
    }{\textcolor{blue}{Table S#3}}%
}

\def\blankpage{%
      \clearpage%
      \thispagestyle{empty}%
      \null%
      \clearpage}

\epstopdfsetup{update} 
\epstopdfsetup{outdir=./figures/}
\epstopdfsetup{suffix=-generated}

%% file: acronymDefinition.tex
\newacronym{se}{SE}{spectroscopic ellipsometry}
\newacronym{apt}{APT}{atom probe tomography}
\newacronym{fib}{FIB}{focused ion beam}
\newacronym{hrxrd}{HRXRD}{high\hyp{}resolution X\hyp{}ray diffraction}
\newacronym{xrd-rsm}{XRD\hyp{}RSM}{X\hyp{}ray diffraction reciprocal space mapping}
\newacronym{rsm}{RSM}{reciprocal space mapping}
\newacronym{fwhm}{FWHM}{full\hyp{}width half maximum}
\newacronym{edx}{EDX}{energy\hyp{}dispersive X\hyp{}ray spectroscopy}
\newacronym{tem}{TEM}{transmission electron microscope}
\newacronym{hrtem}{HRTEM}{High\hyp{}resolution transmission electron microscopy}
\newacronym{stem}{STEM}{scanning transmission electron microscopy}
\newacronym{mocvd}{MOCVD}{metal\hyp{}organic chemical vapor deposition}
\newacronym{cvd}{CVD}{chemical vapor deposition}
\newacronym{sem}{SEM}{scanning electron microscopy}
\newacronym{beol}{BEOL}{back\hyp{}end\hyp{}of\hyp{}line}
\newacronym{cmos}{CMOS}{complementary metal\hyp{}oxide\hyp{}semiconductor}
\newacronym{haadf-stem}{HAADF\hyp{}STEM}{high\hyp{}angle annular dark field scanning transmission electron microscopy}
\newacronym{rms}{RMS}{surface roughness}
\newacronym{rta}{RTA}{rapid thermal annealing}
\newacronym{lta}{LTA}{laser thermal annealing}
\newacronym{ia}{IA}{isothermal annealing}

\newacronym{afm}{AFM}{atomic force microscopy}
\newacronym[\glslongpluralkey={misfit dislocations}]{md}{MD}{misfit dislocation}
\newacronym{td}{TD}{threading dislocation}
\newacronym{fib}{FIB}{focused\hyp{}ion beam}
\newacronym{ctlm}{CTLM}{circular transfer length method}
\newacronym{tlm}{TLM}{transfer length method}
\newacronym{rtse}{RTSE}{room temperature spectroscopic ellipsometry}
\newacronym{vs}{VS}{virtual substrate}
\newacronym{vca}{VCA}{virtual crystal approximation}
\newacronym{pecvd}{PECVD}{enhanced chemical vapor deposition}
\newacronym{aoi}{AOI}{angle of incidence}
\newacronym[\glslongpluralkey={critical\hyp{}points}]{cp}{CP}{critical\hyp{}point}
\newacronym{cppb}{CPPB}{critical point parabolic band}


%% file: 0_abstract.tex
\begin{abstract} 
\medskip
\SiGeSn{}{}{} alloys are promising silicon-compatible semiconductors for monolithic infrared photonics. However, their metastable nature limits the thermal budgets available for post-growth device processing. The mechanisms governing their thermal degradation also remain unresolved. Here, we investigate the thermal stability of \SiGeSn{0.08}{0.88}{0.04} alloys using \textit{in situ} \ac{se} during isothermal annealing at \SI{550}{\degreeCelsius}. We show that adding an ultrathin oxide cap kinetically suppresses \ce{Sn} exchange with the free surface while leaving bulk diffusion pathways largely unaffected. Uncapped films undergo phase separation after ${\sim}\SI{50}{\min}$, accompanied by void formation, a 60\% thickness reduction, and a ${\sim}\SI{400}{\meV}$ blueshift of the $E_2$ \ac{cp} transition, consistent with substitutional \ce{Sn} depletion from the probed volume through surface segregation. In contrast, oxide-capped films exhibit a small compositional change ($<1$~at.\% \ce{Sn}) and optical shift ($<\SI{20}{\meV}$) over the same period, with suppressed void formation, strain relaxation, and alloy decomposition. This surface-kinetic control additionally yields a 25-fold reduction in contact resistivity relative to annealed uncapped alloys. These results identify surface \ce{Sn} transport as the dominant degradation pathway in \SiGeSn{}{}{} and demonstrate that an ultrathin oxide cap extends the thermal stability of metastable group-\RNum{4} alloys, providing a practical route toward their integration into advanced silicon photonic and electronic platforms.

\end{abstract}

%% file: 1_introduction_APL.tex
\par Silicon-germanium-tin (\SiGeSn{}{}{}) ternary alloys offer independent control of both lattice parameter and bandgap energy~\cite{attiaoui2014, wirths2016a, moutanabbir2021, olorunsola2022a, reboud2024}. This orthogonal tunability enables lattice-matched heterostructures with tailored band offsets for silicon-integrated photodetectors~\cite{luo2024}, modulators~\cite{hsieh2021}, lasers~\cite{wirths2015, buca2022, seidel2024a, kim2025}, and photovoltaics~\cite{roucka2010, meyer2025a, daligou2026}. However, the low equilibrium solubility of \ce{Sn} in \ce{Si} and \ce{Ge} ($<$1~at.\%) limits high quality, device-grade integration. Non-equilibrium epitaxy can far exceed this limit, but the resulting metastable films restrict subsequent thermal processing. Elevated temperatures drive \ce{Sn} surface segregation~\cite{zaumseil2018, mukherjee2021, braun2022}, bulk interdiffusion~\cite{vondendriesch2020a}, and dislocation-mediated strain relaxation~\cite{comrie2016, nicolas2020}. These pathways, coupled through three-species dynamics absent in binary \GeSn{}{}~\cite{vondendriesch2020a}, together constrain device fabrication and back-end integration.

\par Prior \textit{ex-situ} studies reported \ce{Sn} surface segregation after annealing~\cite{fournier2014, zaumseil2018, braun2022} and compositional broadening at buried interfaces~\cite{vondendriesch2020a}. Surface \ce{Sn} segregation is thermodynamically favored by reduced surface energy and strain relief, whereas bulk interdiffusion is driven by composition-dependent chemical potential gradients near extended defects and interfaces. The two pathways carry distinct kinetic signatures: surface transport proceeds along dislocations with reduced activation barriers, while bulk diffusion follows slower substitutional exchange. Distinguishing them and resolving incubation and kinetic transients before structural collapse require continuous real\hyp{}time characterization free of model-dependent assumptions. \textit{In situ} \ac{se} addresses this challenge by tracking near-surface electronic structure through \ac{cp} transitions in the pseudo-dielectric function, with sub-monolayer sensitivity~\cite{langereis2009, murphy2026}.

\par Here, we combine \textit{in situ} \ac{se} with controlled surface capping to track \SiGeSn{0.08}{0.88}{0.04} heterostructures during \SI{550}{\degreeCelsius} isothermal annealing, well above the \SI{360}{\degreeCelsius} growth temperature. Uncapped and \SI{3}{\nm} \ce{SiO2}-capped variants of otherwise identical structures isolate the kinetic role of the free surface. The uncapped sample exhibits a \SI{50}{\min} incubation followed by structural collapse, whereas the oxide\hyp{}capped layer remains compositionally and structurally intact over \SI{140}{\min} of annealing under \ce{N2}. Post-anneal characterization confirms that suppressing surface exchange prevents void nucleation, delays strain relaxation, and limits interfacial diffusion, establishing surface \ce{Sn} loss as the rate-limiting step. This stability enables high-temperature contact processing, reducing contact resistivity $25$-fold relative to annealed uncapped alloys.

\par Three samples were prepared from an as-grown \SiGeSn{0.08}{0.88}{0.04}/\GeSn{0.97}{0.03}/\ce{Ge}\hyp{} \ac{vs} heterostructure grown by low-pressure \ac{cvd} at \SI{360}{\degreeCelsius} on \ce{Si}(100) (\fref{fig:fig1_design}{a}): sample~$A$ (as-grown), sample~$B$ (annealed, uncapped), and sample~$C$ (\SI{3}{\nm} \ce{SiO2}-capped, then annealed). Preliminary oxide \textit{ex situ} \ac{se} screening confirmed that both \SI{3}{\nm} \ce{SiO2} (\ac{pecvd}, \SI{300}{\degreeCelsius}) and \SI{5}{\nm} \ce{Al2O3} (atomic layer deposition, \SI{200}{\degreeCelsius}) equally suppress surface \ce{Sn} segregation after \SI{550}{\degreeCelsius} annealing for \SI{30}{\min} (\fref{fig:fig1_design}{b}). \textit{In situ} \ac{se} monitors near-surface electronic structure with submonolayer sensitivity by tracking \ac{cp} transitions in the pseudo-dielectric function~\cite{langereis2009}. High-energy features, such as the $E_2$ transition probe the top $\sim\!\SI{20}{\nm}$~\cite{Aspnes1983}, thereby providing surface selectivity without model-dependent fitting~\cite{murphy2026, bruno1996}. This approach has resolved reaction kinetics during plasma
\vfill
\clearpage

\onecolumngrid 
\adjustimage{width=1.0\textwidth,center,
caption={Experimental design and as-grown \SiGeSn{}{}{} heterostructure baseline. (a) Schematic of the three-sample matrix: sample~$A$ (as-grown reference), sample~$B$ (annealed at \SI{550}{\degreeCelsius} for \SI{140}{\min} without capping), and sample~$C$ (capped with \SI{3}{\nm} \ce{SiO2} before identical annealing). Overlaid \ac{afm} topography maps ($20 \times \SI{20}{\micro\meter\squared}$ scans, color scale -10 to +\SI{30}{\nm} for all maps) show surface morphology for each sample. (b) \textit{Ex situ} \ac{se} screening of oxide effectiveness: imaginary pseudo-dielectric function $\langle\epsilon_2\rangle$ versus photon energy for \SiGeSn{0.08}{0.88}{0.04} samples before (blue) and after (red) \SI{550}{\degreeCelsius}/\SI{140}{\min} annealing with \SI{3}{\nm} \ce{SiO2} (left) or \ce{Al2O3} (right) capping. Black curves show as-grown reference (sample A) without any oxide. (c) Cross-sectional low-magnification TEM image of the as-grown \SiGeSn{0.08}{0.88}{0.04}/\GeSn{0.97}{0.03}/\ce{Ge}-VS stack. (d) \Ac{apt} 3D reconstruction with corresponding \ce{Si} and \ce{Sn} compositional profiles across the multilayer. (e) Asymmetric $(\bar{2}\bar{2}4)$ \ac{rsm} of the as-grown heterostructure; blue stars denote 2D fit-derived peak positions used for lattice constant extraction of each layer in the stack. The $(004)$ $2\theta-\omega$ \ac{hrxrd} scan is overlaid (black curve).},label={fig:fig1_design}, nofloat=figure, vspace=\bigskipamount}{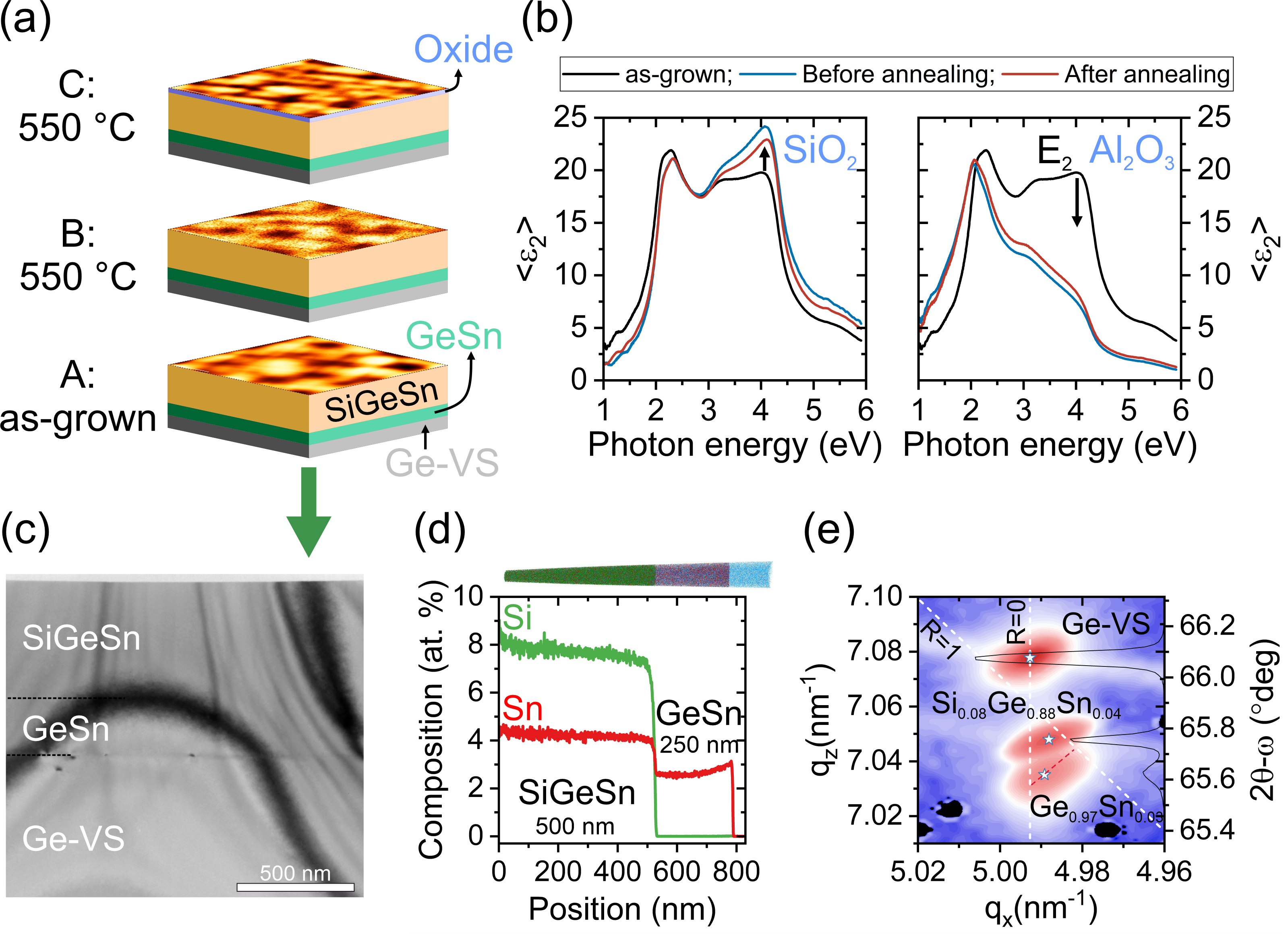} 
\twocolumngrid
\noindent cleaning of \RNum{3}\hyp{}\RNum{5} semiconductors~\cite{losurdo1997}, self-limiting atomic layer deposition~\cite{langereis2009}, and dynamics of oxide formation~\cite{groell2021}.

%% file: 2_body_APL.tex
\par Thermally driven segregation in metastable alloys presents a distinct challenge, as incubation periods and kinetic transients must be captured continuously before the onset of steady-state degradation. Isothermal annealing at \SI{550}{\degreeCelsius} for \SI{140}{\min} was conducted in a Linkam THMS600 stage integrated with a J.A.\ Woollam RC2-XI ellipsometer, acquiring spectra every \SI{10}{\second} at \SI{70}{\degree} incidence over \SIrange{0.7}{6.0}{\eV} under \ce{N2} purge (temperature stability $\pm\SI{1}{\degreeCelsius}$). Additional characterization includes atomic force microscopy (\acs{afm}, $20\,\times\,\SI{20}{\um\squared}$ scans), \ac{apt}, cross-sectional \ac{tem} with \ac{edx}, and asymmetric $(\bar{2}\bar{2}4)$ \ac{xrd-rsm} for strain analysis. Electrical properties were obtained from \ac{tlm} structures with \ce{Ti}/\ce{Au} (5/\SI{60}{\nm}) contacts. More details are provided in the \hl{Supplementary Methods}.

\par The characterization of as-grown \SiGeSn{0.08}{0.88}{0.04} layer (sample~$A$) establishes the baseline for these investigations (\fref{fig:fig1_design}{c--e}). Cross-sectional \ac{tem} reveals \glspl{md} confined to the \GeSn{0.97}{0.03}/\ce{Ge} interface, with no \glspl{td} extending into the \SiGeSn{0.08}{0.88}{0.04} layer at the \ac{tem} imaging scale (\fref{fig:fig1_design}{c}, \hl{Figs.~S1-2}). \Ac{apt} compositional mapping (\fref{fig:fig1_design}{d}) confirms uniform \ce{Si} (8.0~at.\%) and \ce{Sn} (4.0~at.\%)
\twocolumngrid
\adjustimage{width=1.0\textwidth,center,
caption={\textit{In situ} spectroscopic ellipsometry during isothermal annealing at \SI{550}{\degreeCelsius}. (a, b) Pseudo-dielectric function shift $\Delta\langle\epsilon_2\rangle(E,t) = \langle\epsilon_2\rangle(E,t) - \langle\epsilon_2\rangle(E,t_{\text{ref}})$ versus photon energy and elapsed time $(\Delta t=t-t_{\text{ref}})$ for (a) uncapped sample~$B$ and (b) \ce{SiO2}-capped sample~$C$. Color scale encodes magnitude of spectral change relative to the reference state at isothermal onset. (c, d) $E_2$ \ac{cp} parameter shifts versus time: energy $\Delta E_2(t)$ (bottom), broadening $\Delta\Gamma(t)$ (middle), and normalized amplitude $\Delta A(t)/A_0$ (top) for (c) sample~$B$ and (d) sample~$C$. Solid red curves in (c) are logistic fits (\hl{Eq.~S2}); curves in (d) show exponential fits with goodness-of-fit ($R^2 = 0.955$), indicating noise-level fluctuations (\hl{Eq.~S3}).},label={fig:fig2_kinetics}, nofloat=figure, vspace=\bigskipamount}{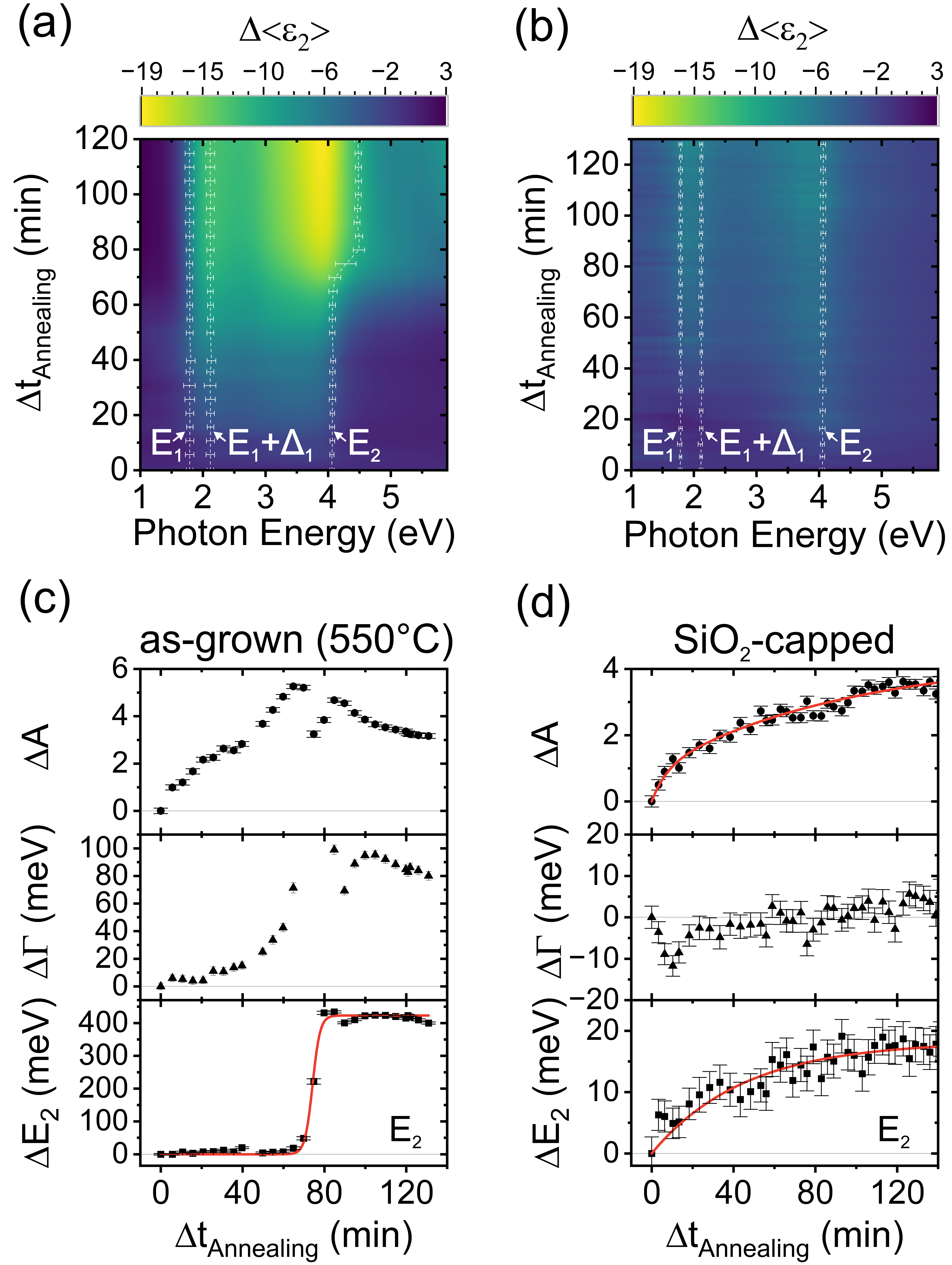}
\twocolumngrid
\begin{table*}[!ht]
  \begin{center}
  \caption{Structural and electrical parameters of \SiGeSn{1-x-y}{x}{y} heterostructures before and after isothermal annealing. Sample A: as-grown. Samples B and C: annealed at \SI{550}{\degreeCelsius} for \SI{140}{\min} under \ce{N2}, without and with \ce{SiO2} capping, respectively. Composition expressed as \ce{Sn}/\ce{Si} atomic ratio ($x/y$) determined by APT (A) or background-corrected TEM-EDX (B, C). \SiGeSn{}{}{} layer thickness $d_{\text{SiGeSn}}$ determined from TEM and APT (A), TEM and EDX (B, C); $1\sigma$ indicate measurement uncertainties ranges. In-plane ($a_{\parallel}$) and out-of-plane ($a_{\perp}$) lattice parameters from asymmetric $(\bar{2}\bar{2}4)$ \ac{xrd-rsm} with \SI{\sim0.002}{\angstrom} combined uncertainty from peak fitting and reciprocal-to-real space conversion. In-plane and out-of-plane strains $\varepsilon_{\parallel}$ and $\varepsilon_{\perp}$ calculated relative to the composition-dependent relaxed lattice constant $a_0$ inferred from the biaxial strain state using elastic constants appropriate to each composition. Contact resistance $R_c$, specific contact resistivity $\rho_c$, and sheet resistance $R_{sh}$ from \ac{tlm} measurements using \ce{Ti}/\ce{Au} (5/\SI{60}{\nm}) contacts.}
  \label{tab:tab1}
  \vspace{2ex}
  \setlength{\tabcolsep}{4pt}
  \renewcommand{\arraystretch}{1.3}
  \begin{tabular}{@{}cccccccccccc@{}}
    \toprule\midrule
    \textbf{Sample} &
    \makecell[c]{\textbf{Surface} \\ \textbf{capping}} &
    \makecell[c]{\textbf{Thermal} \\ \textbf{treatment}} &
    \makecell[c]{\textbf{Sn/Si} \\ \textbf{($x/y$)}} &
    \makecell[c]{\textbf{$d_{\mathrm{SiGeSn}}$} \\ \textbf{(\si{\nm}, $1\sigma$)}} &
    \makecell[c]{\textbf{$\varepsilon_{\parallel}$} \\ \textbf{(\%)}} &
    \makecell[c]{\textbf{$\varepsilon_{\perp}$} \\ \textbf{(\%)}} &
    \makecell[c]{\textbf{$a_{\parallel}$} \\ \textbf{(\AA)}} &
    \makecell[c]{\textbf{$a_{\perp}$} \\ \textbf{(\AA)}} &
    \makecell[c]{\textbf{$R_c$} \\ \textbf{($\Omega$)}} &
    \makecell[c]{\textbf{$\rho_c$} \\ \textbf{($\Omega\cdot$cm$^2$)}} &
    \makecell[c]{\textbf{$R_{sh}$} \\ \textbf{($\Omega/\text{sq}$)}} \\
    \midrule
    $A$ & none & as-grown & $0.50$ & $530 \pm 25$ & $-0.051$ & $+0.038$ & $5.670$ & $5.675$ & $1137$ & $0.190$ & $3886$ \\
    $B$ & none & \makecell{\SI{140}{\min}/\\\SI{550}{\degreeCelsius}} & $0.430$ & $250 \pm 15$ & $-0.025$ & $+0.019$ & $5.674$ & $5.676$ & $1692$ & $0.570$ & $2875$ \\
    $C$ & \makecell{\ce{SiO2} (\SI{3}{\nm}),\\\acs{pecvd}} & \makecell{\SI{140}{\min}\\\SI{550}{\degreeCelsius}} & $0.50$ & $441 \pm 15$ & $-0.048$ & $+0.036$ & $5.676$ & $5.680$ & $431$ & $0.023$ & $4531$ \\
    \midrule\bottomrule
  \end{tabular}
  \end{center}
\end{table*}
\noindent distributions across the $(530\pm25)$~\SI{}{\nm} ternary layer, with compositional gradients below 1~at.\%/$\mu$m. The $(250\pm15)$~\SI{}{\nm} \GeSn{0.97}{0.03} buffer maintains $\sim\!3$~at.\% \ce{Sn}. Asymmetric $(\bar{2}\bar{2}4)$ \ac{xrd-rsm} quantifies the residual biaxial strain state in the \SiGeSn{}{}{} layer: in-plane $\varepsilon_{\parallel} = -(0.051\pm0.002\%)$ (compressive) and out-of-plane $\varepsilon_{\perp} = +(0.038\pm0.002)\%$ (tensile), with lattice constants $a_{\parallel} = (5.670\pm0.002)$~\AA{} and $a_{\perp} = (5.675\pm0.002)$~\AA{} (\fref{fig:fig1_design}{e}, \fref{tab:tab1}{} and S1). Partial relaxation, evidenced by the absence of Pendell\"osung fringes in the $2\theta-\omega$ $(004)$ high-resolution XRD scan, is accommodated by existing \ac{md} arrays without generating new \glspl{td} (\fref{fig:fig1_design}{e}). Surface morphology is smooth, with a root-mean-square roughness of $\sim\!\SI{10}{\nm}$ (\fref{fig:fig1_design}{a}, \hl{Fig.~S3}). This uniform, defect-controlled baseline ensures that thermally activated changes reflect intrinsic material instabilities. The experimental design (\fref{fig:fig1_design}{a}) tests whether bulk degradation requires surface \ce{Sn} loss. The \SI{3}{\nm} \ce{SiO2} layer on sample~$C$ blocks surface exchange while preserving bulk diffusion pathways. If surface loss is rate-limiting, sample~$C$ should remain stable while sample~$B$ degrades; whereas, if bulk interdiffusion dominates, both should degrade similarly. 

\textit{In situ} \ac{se} tracking of the $E_2$ \ac{cp} monitors surface composition during annealing (\fref{fig:fig2_kinetics}{}). The $E_2$ transition probes the top $\sim\!\SI{20}{\nm}$ with compositional sensitivity of $\sim\!\SI{100}{\meV}$/at.\% \ce{Sn}~\cite{Aspnes1983, emminger2020}. \ce{SiO2} capping preserves this spectral feature, whereas \ce{Al2O3} attenuates it via optical interference (\fref{fig:fig1_design}{b}), enabling robust tracking throughout the \SI{140}{\min} cycle. Note that this is the maximum cycle duration the tool permits without annealing pause. The pseudo-dielectric function shift $\Delta\langle\epsilon_2\rangle(E,t) = \langle\epsilon_2\rangle(E,t) - \langle\epsilon_2\rangle(E,t_{\text{ref}})$, referenced to the isothermal onset $t_{\text{ref}}$ at which the sample first reaches \SI{550}{\degreeCelsius}, provides a model-independent fingerprint of near-surface compositional evolution. Sample~$B$ (\fref{fig:fig2_kinetics}{a}) develops strong negative features ($\Delta\langle\epsilon_2\rangle < -19$) between $\sim\!2.5-\SI{3.5}{\eV}$ and positive features below $\sim\!\SI{2}{\eV}$ and above $\sim\!\SI{4}{\eV}$, emerging after $t \approx \SI{50}{\min}$ and saturating by \SI{100}{\min}. This reflects the suppression of \SiGeSn{0.08}{0.88}{0.04} interband transitions and the emergence of \ce{Ge}-richer signatures as \ce{Sn} depletes from the probed volume. Sample~$C$ (\fref{fig:fig2_kinetics}{b}), by contrast, exhibits only small, non-monotonic drifts ($|\Delta\langle\epsilon_2\rangle| \lesssim 7$), confirming compositional stability. Second-derivative spectra $d^2\langle\epsilon_2\rangle/dE^2$ (\hl{Fig.~S4}) resolve the temporal evolution of the $E_1$, $E_1{+}\Delta_1$, and $E_2$ \acp{cp}, corroborating the abrupt onset in sample~$B$ and the stability of sample~$C$. \Ac{cp} parameters (energy $E_2$, broadening $\Gamma$, amplitude $A$) are extracted by fitting the second derivative of $\langle\epsilon\rangle$ to the standard \acf{cppb} lineshape~\cite{jellison1993, fischer2017, emminger2020, imbrenda2021}:
\begin{equation}
  \frac{d^2\epsilon}{dE^2} = Ae^{i\phi}\left(E - E_2 + i\Gamma\right)^{n-2},
  \label{eq:cppb_inline}
\end{equation}
\noindent where $A$ is the oscillator strength, $\phi$ accounts for phase mixing, and $n=-1/2$ corresponds to the three-dimensional $M_0$ \ac{cp} character of $E_2$~\cite{Yu2010}. This fixed exponent enables consistent tracking despite lineshape distortions from compositional inhomogeneity (\hl{Supplementary Methods}). Time-dependent shifts $\Delta E_2(t) = E_2(t) - E_2(t_{\text{ref}})$, $\Delta\Gamma(t) = \Gamma(t) - \Gamma(t_{\text{ref}})$, and $\Delta A(t) = A(t) - A(t_{\text{ref}})$, referenced to the isothermal onset, quantify compositional redistribution, structural disorder, and electronic structure changes, respectively.

\par Sample~$B$ (uncapped) demonstrates a three-phase $E_2$ evolution (\fref{fig:fig2_kinetics}{c}). After a \SI{50}{\min} incubation during which $\Delta E_2$ remains within $\pm\SI{20}{\meV}$, the \ac{cp} blueshifts rapidly, saturating at \SI{423}{\meV} by $t \approx \SI{100}{\min}$. This $S$-shaped trajectory is well-described by a logistic function ($R^2 = 0.996$), with a midpoint $t_0 = (74.2 \pm 0.5)$~min marking the degradation onset and a width $w = (1.86 \pm 0.15)$~min defining the collapse timescale. At the isothermal onset ($t_{\text{ref}}$, \SI{550}{\degreeCelsius}), thermal renormalization has already redshifted $E_2$ by $\approx-\SI{138}{\meV}$ relative to room temperature~\cite{fernando2017, emminger2020}. Concurrent broadening $\Delta\Gamma/\Gamma_0 = +30\%$ exceeds the $\sim\!22\%$ anticipated thermal contribution at \SI{550}{\degreeCelsius}~\cite{fernando2017}, suggesting additional disorder from compositional inhomogeneity. The normalized amplitude drops by $\Delta A/A_0 \approx -15\%$, consistent with reduced oscillator strength as the film transforms into a nanoscale mixture of \ce{Sn}-depleted regions, voids, and oxidized zones. The prolonged incubation and abrupt \SI{1.86}{\min} collapse are characteristic of a threshold-activated process in which accumulated supersaturation overcomes kinetic barriers, triggering autocatalytic transformation. Sample~$C$ (capped) shows no measurable evolution (\fref{fig:fig2_kinetics}{d}). $\Delta E_2$ fluctuates within $\pm\SI{20}{\meV}$ over the full \SI{140}{\min} without monotonic drift. The exponential fit yields an amplitude of \SI{18}{\meV}, but $R^2 = 0.955$ indicates measurement noise rather than systematic kinetics. Broadening remains constant ($|\Delta\Gamma/\Gamma_0| < 5\%$, \hl{Supplementary Section~S2.1}), confirming preserved crystalline order, with the final $E_2$ energy recovering to its room-temperature value. This kinetic contrast identifies surface-mediated \ce{Sn} segregation as the pathway driving bulk degradation at \SI{550}{\degreeCelsius}. The \SI{3}{\nm} \ce{SiO2} layer is sufficient to kinetically suppress \ce{Sn} transport to the free surface while retaining bulk diffusion.

\par Post-anneal structural analysis confirms that capping suppresses all degradation mechanisms (\fref{fig:fig3_tem}{}). Cross-sectional \ac{tem} of the uncapped sample~$B$ reveals void formation throughout the upper \SI{330}{\nm} (\fref{fig:fig3_tem}{a}), with only a $(200\pm30)~\si{nm}$ remnant surviving beneath the degraded region, representing $\sim\!60\%$ thickness loss relative to sample~$A$. \Ac{edx} detects oxygen within these voids (\hl{Fig.~S5}), consistent with secondary oxidation along pathways opened by \ce{Sn} loss. High-magnification \ac{tem} resolves filamentary \ce{Sn}-enriched regions extending inward from the surface, confirming directional outward \ce{Sn} transport (\hl{Fig.~S5}). \Ac{edx} line profiles reveal heterogeneous redistribution. Scan $1\to1'$ shows \ce{Sn} dropping from 4~at.\% to near zero, with \ce{O} reaching 40~at.\%, while scan $2\to2'$ exhibits surface depletion over the top \SI{100}{\nm} but near-baseline \ce{Sn} below, establishing that degradation nucleates at discrete sites rather than propagating uniformly (\hl{Fig.~S5}). Conversely, cross-sectional \ac{tem} of the capped sample~$C$ (\fref{fig:fig3_tem}{b}) shows a fully intact \SiGeSn{}{}{} layer without voids, with composite \ac{edx} maps highlighting uniform \ce{Sn} and no detectable oxygen within the ternary film. \Ac{afm} corroborates this contrast: sample~$B$ reaches an \ac{rms} roughness of $(32\pm4)~\si{\nm}$ compared to the as-grown $(10.1\pm0.5)~\si{\nm}$, while sample~$C$ remains at $(10.5\pm0.3)~\si{\nm}$, indistinguishable from sample~$A$ (\fref{fig:fig1_design}{a}, \hl{Fig.~S3}).

\par \Ac{xrd-rsm} confirms the compositional redistribution (\fref{fig:fig3_tem}{c,d}). Sample~$B$ (\fref{fig:fig3_tem}{c}) develops a new diffraction peak at $Q_z \approx \SI{7.06}{\per\nm}$, identifying a \GeSn{0.99}{0.01} surface segregation layer absent in both the as-grown state (\fref{fig:fig1_design}{e}) and sample~$C$ (\fref{fig:fig3_tem}{d}). Its remnant \SiGeSn{}{}{} layer yields $\varepsilon_{\parallel} = -(0.025\pm0.002)$\% ($a_{\parallel} = 5.674\pm\SI{0.002}{\angstrom}$, $a_{\perp} = 5.676\pm\SI{0.002}{\angstrom}$, \fref{tab:tab1}{}), a $+0.026$\% shift relative to sample~$A$, with both the \SiGeSn{}{}{} and \GeSn{}{} peaks migrating toward the full-relaxation line ($R=1$, open red stars in \fref{fig:fig3_tem}{c}). This partial relaxation reflects thermally activated glide of existing \ac{md} arrays at the \GeSn{0.97}{0.03}/\ce{Ge}\hyp{}VS interface, relieving strain without nucleating new \glspl{td}, as reported for \ce{SiGe}/\ce{Si} systems~\cite{legoues1991, legoues1992, tersoff1994}. The absence of observable new threading dislocations in sample~$B$ supports this mechanism (\hl{Fig.~S2}). In contrast, sample~$C$ retains $\varepsilon_{\parallel} = -(0.048\pm0.002)$\% (\fref{fig:fig3_tem}{d}), within $+0.003$\% of the as-grown state ($-0.051\pm0.002$\%, \fref{tab:tab1}{}), which is comparable to the thermal expansion effect~\cite{fernando2017}, confirming that oxide passivation limits structural evolution. 

\par The structural data resolve a central question in \SiGeSn{}{}{} thermal stability: does blocking surface transport sup- 
\onecolumngrid
\adjustimage{width=0.9\textwidth,center,
caption={Post-anneal structural validation of surface-exchange suppression. (a) Top: Cross-sectional \ac{tem} of uncapped sample~$B$ after \SI{140}{\min} at \SI{550}{\degreeCelsius}. Bottom: composite \ac{edx} map showing spatial correlation between voids (dark regions) and oxygen (green), with \ac{edx} line scans 1$\to$1$'$ and 2$\to$2$'$ quantifying spatially heterogeneous compositional redistribution (\hl{Fig.~S5}). (b) Left: cross-sectional \ac{tem} of \ce{SiO2}-capped sample~$C$ after identical thermal treatment. Right: \ac{edx} elemental maps of \ce{Sn} (red) and \ce{O} (green) confirm preserved \ce{Sn} and absence of \ce{O} in the \SiGeSn{}{}{} layer. Asymmetric $(\bar{2}\bar{2}4)$ \acp{rsm} of (c) uncapped sample~$B$ and (d) \ce{SiO2}-passivated sample~$C$. Blue stars mark as-grown peak positions (from \fref{fig:fig1_design}{e}); red stars mark post-anneal positions. White dashed lines indicate full relaxation ($R=1$) and pseudomorphic strain ($R=0$) trajectories. In panel (c), the red star at $Q_z \approx \SI{7.06}{\per\nm}$ identifies the \GeSn{0.99}{0.01} surface segregation layer, which is absent in panel (d). The log-scale color bar encodes diffracted intensity. Structural parameters are in \fref{tab:tab1}{}.},label={fig:fig3_tem}, nofloat=figure, vspace=\bigskipamount}{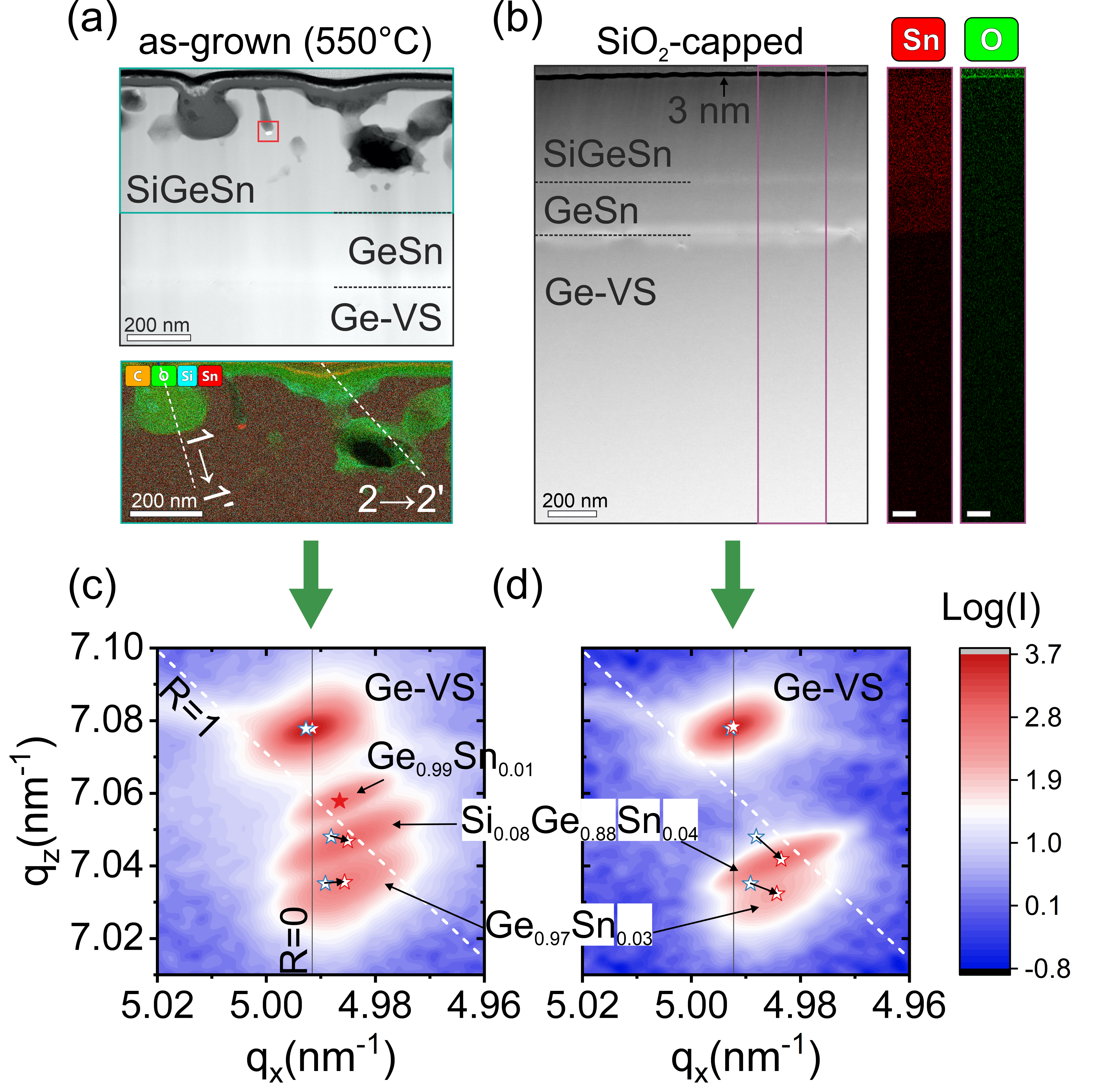}
\twocolumngrid
\noindent press all three degradation paths simultaneously? Void nucleation ($\sim\!60\%$ thickness loss in sample~$B$ versus none in sample~$C$), strain relaxation ($\Delta\varepsilon_{\parallel} = +0.023$\% versus $+0.003$\%), and formation of a \GeSn{0.99}{0.01} segregation layer present only in sample~$B$. establishing a clear surface-kinetic hierarchy. The $+0.023$\% strain contrast is five times the thermal expansion baseline ($\sim\!0.004$\%, \hl{Supplementary Section 2}~\cite{reeber1996,roucka2010}), reflecting composition-driven \glspl{md} glide. \ce{Sn} depletion in the upper \SI{330}{\nm} lowers the local lattice constant and generates stress gradients at the voided/remnant interface that drive dislocation motion~\cite{hull1992}, while compositional gradients between the two regions sustain local interdiffusion, coupling surface \ce{Sn} segregation to bulk structural changes. If bulk interdiffusion proceeded independently, capping would only delay degradation, and stabilization would require compositional grading or strain engineering. The complete suppression observed in sample~$C$ over \SI{140}{\min} rules this out, demonstrating that a nanoscale interfacial barrier alone can thermally stabilize the alloy.

\par Surface-mediated \ce{Sn} transport is driven by three thermodynamic penalties. First, \ce{Sn} exceeds its equilibrium solubility in both \ce{Ge}($<$1~at.\%)~\cite{trumbore1956} and \ce{Si} ($<$0.42~at.\%)~\cite{trumbore1960, olesinski1984}, generating a chemical potential gradient $\Delta\mu \approx k_BT\ln(x/x_{\text{eq}}) \approx \SI{35}{\meV}$ per excess atom toward the surface. Second, the lower surface energy of \ce{Sn} (\SI{0.56}{\joule\per\meter\squared}) relative to \ce{Ge} (\SI{0.61}{\joule\per\meter\squared})~\cite{barnett1985} favors surface occupancy and segregation. Third, the larger covalent radius of \ce{Sn} ($\sim\!\SI{1.40}{\angstrom}$) compared to \ce{Ge} ($\sim\!\SI{1.22}{\angstrom}$)~\cite{pyykko2009} stores elastic energy in the compressively strained layer ($\varepsilon_{\parallel} = -0.051$\%), rendering migration to the free surface an efficient strain relief pathway. Bulk substitutional diffusion cannot transport \ce{Sn} to the surface on the observed timescale: with bulk activation enthalpy $E_a^{\text{Ge}} \approx 2.6$--$\SI{2.7}{\eV}$~\cite{brotzmann2008}, the \ce{Sn} diffusion length through the alloy is only $L \sim \sqrt{Dt} \approx \SI{0.3}{\nm}$ over \SI{50}{\min}, three orders of magnitude below the \SI{530}{\nm} film. Instead, the pre-existing misfit dislocation network at the \GeSn{0.97}{0.03}/\ce{Ge}\hyp{}\ac{vs} interface provides fast-diffusion conduits with reduced barriers ($E_a^{\text{pipe}} \approx 1.5$--$\SI{1.8}{\eV}$), enabling pipe diffusion along dislocation cores and their connected threading arms to the surface. This is consistent with prior \GeSn{}{} studies, where segregation occurs once the misfit dislocation line density exceeds $\sim\!5\times10^5$~cm$^{-1}$~\cite{stanchu2021}, and accounts for the filamentary \ce{Sn}-rich morphology in \ac{tem} (\hl{Fig.~S5(a)}). The \SI{50}{\min} incubation reflects the time required to build critical \ce{Sn} supersaturation at dislocation outcrops. Void nucleation follows Kirkendall dynamics~\cite{wang2013}: sustained outward \ce{Sn} flux along dislocation cores outpaces the compensating inward matrix flux, creating vacancy supersaturation. Once voids nucleate, they create additional fast-diffusion paths and relieve local stress, driving the autocatalytic collapse within \SI{1.86}{\min} (\fref{fig:fig2_kinetics}{c}).

\par The \SI{3}{\nm} \ce{SiO2} layer suppresses degradation by blocking all \ce{Sn} pathways to the free surface: dislocation-mediated pipe diffusion, interstitial migration, and substitutional exchange at the \SiGeSn{}{}{}/\ce{SiO2} interface. Any \ce{Sn} reaching this interface must traverse the amorphous oxide. Accommodating oversized \ce{Sn} atoms in \ce{SiO2} imposes a migration barrier far exceeding that in the crystalline alloy ($E_a^{\text{oxide}} \gg E_a^{\text{alloy}}$), reducing \ce{Sn} diffusion path within the oxide to $\sim\!\SI{0.1}{\nm}$ over \SI{140}{\min} at \SI{550}{\degreeCelsius}~\cite{mcbrayer1986}, two orders of magnitude below the oxide thickness. Eliminating the free-surface sink removes the driving force for net \ce{Sn} transport: with no irreversible loss pathway, no chemical-potential gradient develops, and \ce{Sn} remains substitutionally uniform throughout the ternary film, consistent with the uniform \ce{Sn} profile in sample~$C$ (\fref{fig:fig3_tem}{b}). In uncapped film  (sample~$B$), the free surface maintains a continuous flux by irreversibly removing \ce{Sn} through surface oxide formation and droplet segregation, thereby sustaining the gradient that drives vacancy supersaturation and autocatalytic void growth.

\par Contact integration in \SiGeSn{}{}{} heterostructures has been constrained by empirical thermal ceilings that depend on alloy composition, strain state, and growth temperature, typically falling in the \SIrange{300}{400}{\degreeCelsius} range for \ce{Sn}-rich alloys. Nickel-stano-germanide (\ce{NiGeSn}) contacts on \GeSn{}{} achieve low resistivity at \SI{325}{\degreeCelsius}, but degrade above \SI{400}{\degreeCelsius} through \ce{Sn} out-diffusion~\cite{schulte2017}, precluding sequential backend steps such as dopant activation (\SIrange{500}{700}{\degreeCelsius}), \ce{Ti}/\ce{TiN} barrier anneals, and \ac{pecvd} dielectrics (\SIrange{400}{450}{\degreeCelsius}). Prior approaches mitigated rather than removed this constraint, lowering metal-gate deposition temperatures, merging germano-stannide formation with dopant segregation, or restricting backend flows at $<\SI{350}{\degreeCelsius}$~\cite{schulte2016}, without establishing whether the ceiling is intrinsic or an artifact of unpassivated surface kinetics. Our controlled comparison in \fref{fig:fig3_tem}{} demonstrates that blocking surface \ce{Sn} exchange before thermal treatment decouples contact optimization from compositional degradation. This extends the processing window to \SI{550}{\degreeCelsius} while preserving the composition and strain required for optimal optoelectronic performance. The \ac{tlm} measurements below validate this approach (\hl{Supplementary Methods}).

\par This structural stability directly enhances electrical performance (\fref{fig:fig4_electrical}{}). All samples underwent \ce{HF} oxide removal, \ce{Ti}/\ce{Au} (\SI{5}/\SI{60}{\nm}) metallization, and a \SI{380}{\degreeCelsius}/\SI{30}{\second} post-contact \acf{rta} in forming gas. \ac{tlm} measurements with gap spacings $d_s = \SIrange{11}{77}{\micro\meter}$ quantify how pre-metallization annealing affects contact resistivity $\rho_c$ (\fref{tab:tab1}{}). Sample~$A$ shows $\rho_c = \SI{0.190}{\ohm\cm\squared}$, typical for \ce{Ti} on $p$-type \ce{Ge}-rich alloys, where \SI{380}{\degreeCelsius} is insufficient for optimal \ce{Ti}\hyp{}\ce{Ge} interfacial layer formation. Pre-metallization annealing of sample~$B$ at \SI{550}{\degreeCelsius} degrades contact despite subsequent \ce{HF} cleaning and \ac{rta}: $\rho_c$ triples to \SI{0.570}{\ohm\cm\squared}, with non-linear, asymmetric $I$--$V$ curves (\fref{fig:fig4_electrical}{a}, blue). Void networks in the \SI{330}{\nm} degraded layer disrupt current percolation and reduce the effective contact area. Severe surface roughness (\ac{rms} = \SI{32}{\nm}) prevents reliable planar contact formation, so $\rho_c$ reflects a degraded, non-ideal interface rather than an intrinsic contact resistivity. Spatially non-uniform \ce{Sn} depletion exacerbates the \ce{Ti}–\ce{Ge} interfacial layer by creating locally varying Schottky barriers, traps, and Fermi level pinning. The sheet resistance drops to $R_{\text{sh}} = \SI{2875}{\ohm}/\square$, reflecting the \SI{60}{\percent} thickness loss rather than conductivity improvement.

\par Sample~$C$ behaves inversely: \SI{550}{\degreeCelsius} annealing preserves composition and crystallinity while conditioning the surface, removing hydrocarbons and annihilating growth-related point defects. Subsequent oxide stripping and \SI{380}{\degreeCelsius} \ac{rta} reduce $\rho_c$ to \SI{0.023}{\ohm\cm\squared} with a linear, symmetric $I$--$V$ response (\fref{fig:fig4_electrical}{a}, orange), confirming ideal ohmic contacts. The preserved \ce{Sn} content ($\sim\!4$~at.\%) maintains the narrow bandgap and high valence band edge, supporting efficient hole injection, while the compositional uniformity promotes a coherent $\sim\!\SI{3}{\nm}$ \ce{Ti}\hyp{}\ce{Ge} interfacial layer. This layer reduces interface trap density and supports field-assisted tunneling~\cite{atalla2023}. The \ac{tlm} ana-
\onecolumngrid
\adjustimage{width=1\textwidth,center,
caption={Electrical characterization of \SiGeSn{}{}{} heterostructures before and after isothermal annealing at \SI{550}{\degreeCelsius}. (a) Current--voltage ($I$--$V$) characteristics across a $d_s = \SI{77.5}{\um}$ contact gap for as-grown sample~$A$ (purple), unpassivated sample~$B$ (blue), and \ce{SiO2}-passivated sample~$C$ (orange).  (b) Total resistance $R_T$ versus contact spacing $d_s$ from \ac{tlm} measurements. Solid lines are linear fits extracting contact resistivity $\rho_c$ and sheet resistance $R_{sh}$ (\fref{tab:tab1}{}). Inset: optical micrograph of the \ac{tlm} test structure with \ce{Ti}/\ce{Au} contact pads and variable gap spacings.},label={fig:fig4_electrical}, nofloat=figure, vspace=\bigskipamount}{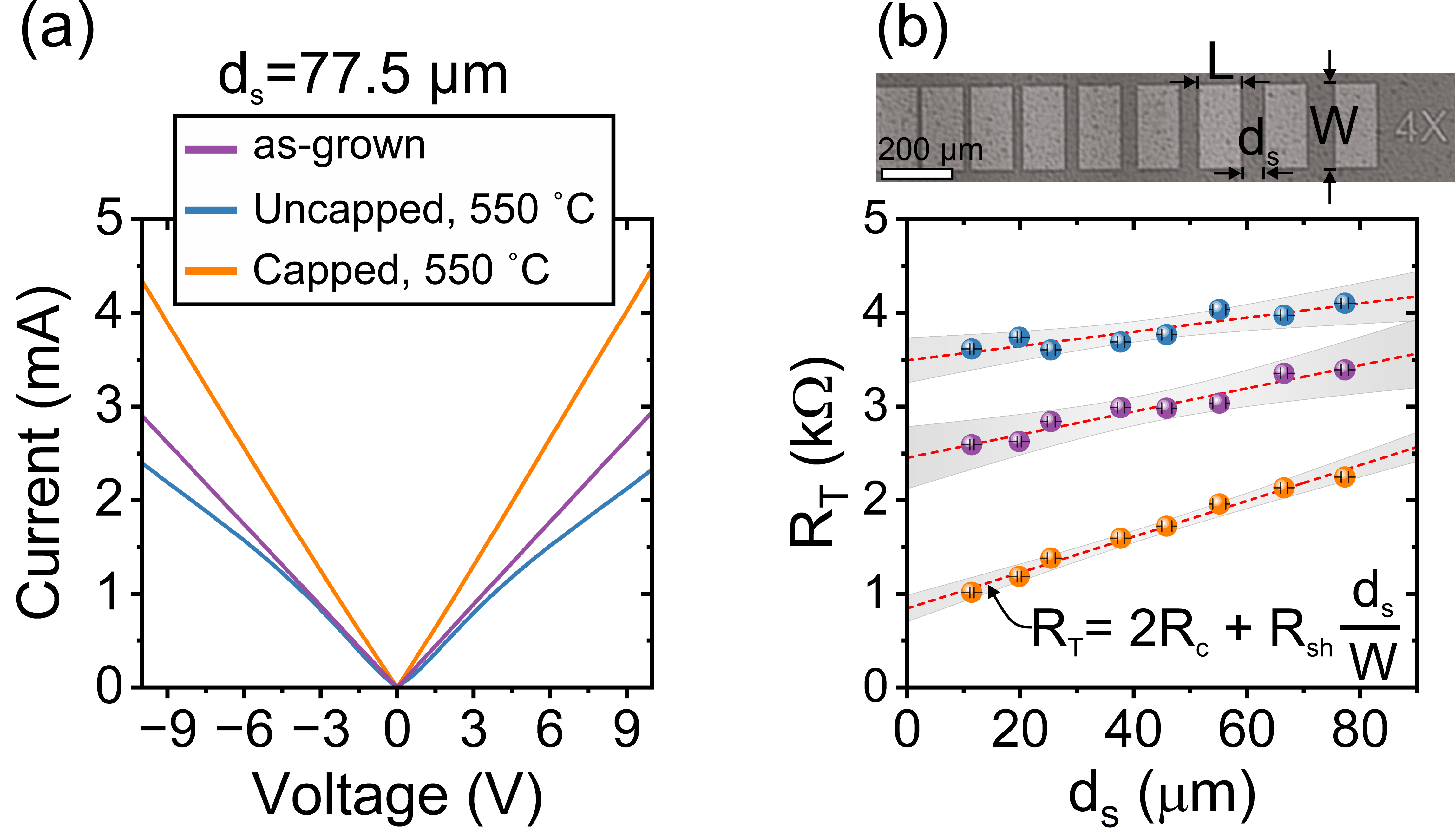}
\twocolumngrid
\noindent lysis (\fref{fig:fig4_electrical}{b}) yields sheet resistance $R_{\text{sh}} = \SI{4531}{\ohm}/\square$, consistent with restored semiconducting transport. The \SI{17}{\percent} increase over sample~$A$ stems from the smaller thickness (\SI{441}{\nm} vs. \SI{530}{\nm}, \fref{tab:tab1}{}) via $R_{\text{sh}} \propto 1/t$. Crucially, $\rho_c = \SI{0.023}{\ohm\cm\squared}$ is \num{8} times lower than the as-grown sample~$A$ and \num{25} times lower than in sample~$B$, indicating that passivation-enabled thermal processing improves contact quality without compromising compositional integrity. This \SI{550}{\degreeCelsius} stability offers a \SIrange{50}{150}{\degreeCelsius} thermal margin above typical \ac{beol} limits (\SIrange{400}{500}{\degreeCelsius}), enabling \SiGeSn{}{}{} integration within foundry-compatible thermal budgets.

%% file: 3_conclusion.tex
\par In summary, ultrathin \ce{SiO2} passivation kinetically blocks \ce{Sn} transport to the free surface, stabilizing metastable \SiGeSn{0.08}{0.88}{0.04} heterostructures during \SI{550}{\degreeCelsius} isothermal annealing for \SI{140}{\min}. \textit{In situ} \acf{se} reveals that uncapped films undergo a \SI{423}{\meV} $E_2$ blueshift after a \SI{50}{\min} incubation, with structural collapse completing within \SI{1.86}{\min}. Oxide-capped films show no detectable compositional drift ($|\Delta E_2| < \SI{20}{\meV}$), strain evolution ($|\Delta\varepsilon_{\parallel}| < 0.003\%$), or linewidth broadening ($|\Delta\Gamma/\Gamma_0| < 5\%$) over the same exposure. Post-anneal characterization confirms that suppressing surface exchange arrests void nucleation ($\sim\!60\%$ thickness loss in uncapped \textit{vs.} none in \ce{SiO2}\hyp{}capped), strain relaxation ($51\%$ \textit{vs.} $<1\%$), and formation of a \GeSn{0.99}{0.01} segregation layer present only in uncapped films. This surface-kinetic control lowers contact resistivity to \SI{0.023}{\ohm\cm\squared}, \num{8} times below the as-grown value, providing \SIrange{50}{150}{\degreeCelsius} of thermal margin above standard \ac{beol} temperatures and enabling \SiGeSn{}{}{} integration within foundry-compatible windows. The underlying mechanism, in which nanoscale interfacial barriers decouple free-surface thermodynamics from bulk kinetics, extends beyond planar heterostructures: ultrathin \ce{Al2O3} passivation similarly stabilizes \ce{Ge}/\GeSn{}{} core/shell nanowires at \SI{450} {\degreeCelsius}~\cite{attiaoui2026}, establishing oxide-mediated kinetic suppression as a general strategy against surface-initiated degradation in metastable semiconductor alloys.

%% file: acknowledgments.tex
\medskip
\par See the supplementary materials for experimental details, additional characterization, and \ac{se} spectral fitting.

\medskip
\par The authors thank Faqrul A. Chowdhury, Mahmoud R. M. Atalla, Patrick Daoust for fruitful discussions, Joel Bouchard for support with the CVD system, F.A. Chowdhury for assistance with the electrical measurements, and Bill Baloukas for their help with the spectroscopic ellipsometry measurements. We are grateful to J\'er\^ome Nicolas for his valuable support with the XRD-RSM measurements. O.M. acknowledges support from NSERC Canada, Canada Research Chairs, Canada Foundation for Innovation, and Defence Canada (Innovation for Defence Excellence and Security, IDEaS).\\